\documentclass[letterpaper]{jpconf}
\usepackage{graphicx}
\begin{document}
\title{Weak interactions and quasi-stable particle energy loss}

\author{M. H. Reno$^1$, I. Sarcevic$^2$ and J. Uscinski$^2$}

\address{$^1$Department of Physics and Astronomy, University of Iowa, Iowa City, Iowa 52242\\
$^2$Department of Physics, University of Arizona, Tucson, Arizona}

\ead{mary-hall-reno@uiowa.edu, ina@physics.arizona.edu,
uscinski@physics.arizona.edu}

\begin{abstract}
We discuss the interplay between electromagnetic energy loss and weak interactions in the context of quasistable particle particle propagation through materials. As specific examples, we consider staus, where weak interactions may play a role, and taus, where they don't.
\end{abstract}

\section{Introduction}

Large underground and under-ice neutrino telescopes detect through-going charged particles by the
Cherenkov light they emit in transit. Proposals to measure tau neutrino fluxes, e.g.,
$\nu_\tau\rightarrow \tau$ followed by $\tau$ decay (via the double bang signal [1] 
or either the tau production or decay vertex) 
are
possible in the 1-10 PeV neutrino energy range [2]. 
For higher energies, above about
20 PeV, taus may
be considered quasi-stable particles, in that their decay lengths are larger than the size of
the detector. Muons are quasi-stable particles on the scale of 1 km even for energies in the GeV range.

Recently, supersymmetry models with weak scale supersymmetry breaking [3] have led to studies of
quasi-stable stau ($\tilde{\tau}$) signals in neutrino telescopes [4-7].
In these studies, pairs of staus come from the decays of supersymmetric particles produced by
neutrinos interacting in the Earth. The signature would be a pair of tracks from charged particles,
the staus, passing through the detector. Neutrino production of supersymmetric particles is suppressed, but
the signal can be enhanced by the large effective volume of the detector, characterised by the cross sectional
area of the detector multiplied by the range of the staus.

Electromagnetic energy loss is a key feature of charged particle ranges
for muons, high energy taus and high energy
staus. On average, the energy loss
per unit distance is
\begin{equation}
\Biggl\langle \frac{dE}{dX}\Biggr\rangle =
-(\alpha + \beta E)
\end{equation}
where $\alpha \simeq 3\times 10^{-3}$ GeV cm$^2$/g is due to ionization energy loss, while $\beta$
depends on the mass of the traveling particle with contributions from bremsstrahlung,
$e^+e^-$ pair production and photonuclear interactions.
For muons, $\beta_\mu\simeq 3\times 10^{-6}$ cm$^2$/g, while for taus, $\beta_\tau\simeq 2\times 10^{-7}$ cm$^2$/g.
Schematically, for staus,
\begin{equation}
\beta_{\tilde{\tau}}\sim \Biggl(\frac{m_\tau}{m_{\tilde{\tau}}}\Biggr) \ 2\times 10^{-7}\ {\rm cm}^2/{\rm g}\ .
\end{equation}
We have evaluated the energy loss parameter $\beta$ for staus more quantitatively in Ref. [7].

The parameter $\beta$ 
can be converted to a distance scale by
multiplying by a density. For the figures below, the
density $\rho=2.65$ g/cm$^3$ is for standard rock. 
There is a weak energy dependence for the $\beta$ parameters for each of the particles. For taus,
$(\beta_\tau \rho)^{-1}\simeq 10$ km at a PeV and decreases with energy.
Muons have a characteristic distance of $(\beta_\mu\rho )^{-1}\simeq 1$ km. Staus, on the other hand,
have $(\beta\rho)^{-1}\simeq 1000$ km for $E=1$ PeV. 

Another characteristic distance relevant to muon, tau and stau propagation through ice, water or rock is
the weak interaction length. For $E_\tau=10^{10}$ GeV, the tau charged-current interaction length is
of order $10^3$ km. The muon and stau interaction lengths are of the same order of magnitude.
By comparing to $(\beta\rho)^{-1}$, we see that weak interactions are potentially important for staus, but not
for taus. For muons, the distance scales are even further separated,
so we confine our discussion to taus and staus.  

In the next section, we discuss the weak interaction contributions to stau energy loss in the Earth.
We have evaluated the range of taus and staus, for specific SUSY model
parameter choices, using a Monte Carlo simulation including stochastic energy loss. 
In the last section, we compare the stau ranges with and without including weak interactions based on the Monte Carlo simulation results.

\section{Weak interactions of staus}

\begin{figure}[h]
\begin{minipage}{18pc}
\includegraphics[width=14pc,angle=270]{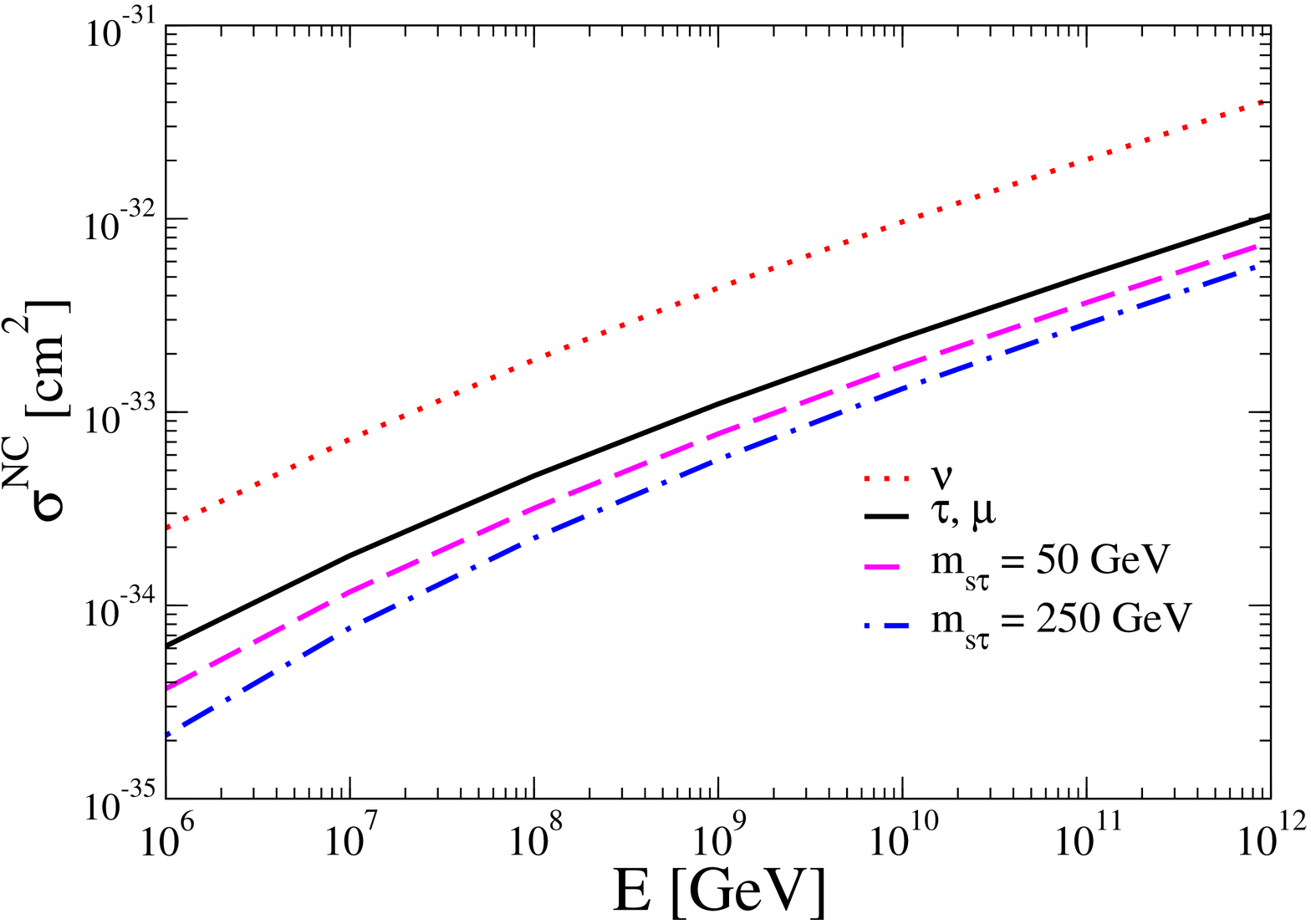}
\caption{\label{label}Neutral current cross section versus
energy for neutrinos, muons, taus and staus with
$m_{\tilde{\tau}}=50,\ 250$ GeV and $\cos2\theta_f=1$.}
\end{minipage}\hspace{2pc}%
\begin{minipage}{18pc}
\includegraphics[width=14pc,angle=270]{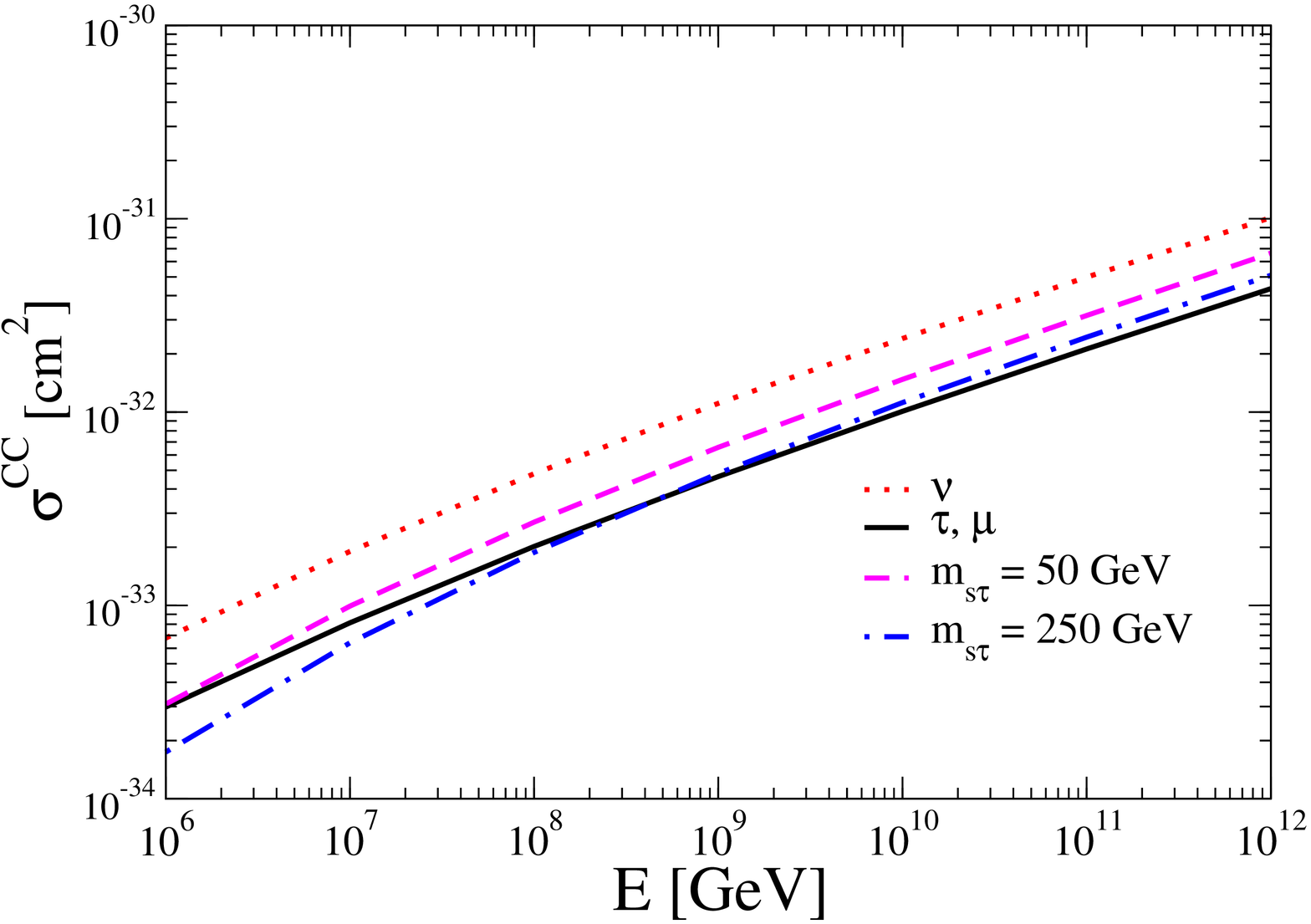}
\caption{\label{label}The charged current cross section versus energy.
For staus, the curves are shown for $\sin\theta_f=1$.}
\end{minipage} 
\end{figure}

The charged current (CC) and neutral current (NC) 
interactions of taus are related to neutrino interactions,
with the appropriate substitution of couplings and the tau spin
average.
Stau weak interaction cross sections depend on the stau mass,
the relation between
the stau mass eigenstates and the partners of the left-handed and
right-handed taus and the scalar couplings with the $W$ and
$Z$. In these models, the next-to-lightest supersymmetric
particle is the stau with
$$\tilde{\tau}=\cos\theta_f\,\tilde{\tau}_R 
+\sin\theta_f\,\tilde{\tau}_L\ .
$$
We take $\theta_f$ to be a free parameter. For both 
NC and CC stau interactions, the couplings 
depend on $\theta_f$. 

In Fig. 1, we show the NC cross section for neutrinos, taus, muons
and staus ($m_{\tilde{\tau}}=50,\ 250$ GeV) 
picking $\cos2\theta_f=1$. We show this range of masses to give an
indication of the dependence on the stau mass. LEP experiments
constrain the stau mass to be larger than $\sim 100 $ GeV [8].
Fig. 2 shows the CC cross section for
the same incident particles, this time for staus with $\sin\theta_f=1$,
the maximal value. If $\tilde{\tau}=\tilde{\tau}_R$, the CC cross
section for staus vanishes.

Electromagnetic and NC interactions act the same way, in that they 
reduce the initial energy of the incident particle to a lower energy. 
For taus, and even for staus, the NC energy loss is small
compared to the electromagnetic energy loss. In what follows we
neglect the NC interactions.
The CC interaction, however, is important for staus if $\sin\theta_f=1$.
The CC interaction removes staus from the incident flux.
For taus, the CC interaction is not important compared
to the electromagnetic energy loss, as we show in
the next section.

\section{Ranges}

In this section, we show results for the range for taus and staus and the effect of
CC interactions [9]. For taus, one approach has been to look at the different scales:
the decay length, $1/(\beta\rho)$ and the charged current interaction length
$L_\tau^{CC}$ [10]. Here, instead, we show the results for a full stochastic
treatment of tau energy loss including weak interactions [11,12]. The range for taus
in rock is shown with the solid lines in Fig. 3. The dashed lines show the characteristic
scales. With the exception of the decay length for low energies, none of the characteristic
scales aligns with the range. For electromagnetic energy loss, this is not surprising
since $1/(\beta\rho)$ is not a solution to Eq. (1). The CC interaction length is
long compared to the tau range determined from electromagnetic interactions alone, so
its effect is not visible on the scale of this figure.

Fig. 4 shows the characteristic scales (dashed lines) for a stau traveling through
rock. The stau mass is 150 GeV and the decay parameter is $\sqrt{F}=10^7$ GeV for
this figure, where
\begin{equation}
c\tau = \Biggl( \frac{\sqrt{F}}{10^7\ {\rm GeV}}\Biggr)^4
\Biggl( \frac{100\ {\rm GeV}}{m_{\tilde{\tau}}}\Biggr)^5\ 
10\ {\rm km}.
\end{equation}
For this choice of $\sqrt{F}$, ionization energy loss characterized
by the distance scale $E/\alpha$ is relevant while $Ec\tau/m_{\tilde{\tau}}$ is not.
The minimum energy of the stau 
for evaluating the range is $E_0=10^3$ GeV in Fig. 4.
The upper solid line shows the range when $\sin\theta_f=0$, namely, no CC interactions
affect the range. The lower solid line shows the effect of including maximal
CC interactions. Similar results are obtained for other masses [9].
We note again in Fig. 4 that the characteristic scales at high energies
are only qualitative representations of the stau range.

\begin{figure}[h]
\begin{minipage}{18pc}
\includegraphics[width=14pc,angle=270]{fig3.ps}
\caption{\label{label}Characteristic distances for taus in rock: dashed
lines for decay length ($\gamma c \tau$), $1/(\beta\rho)$ and
$L_\tau^{CC}=1/(N\sigma_{CC}\rho)$, and the solid line for
the tau range. The curves with and without CC interactions are indistinguishable on the scale of this figure.}
\end{minipage}\hspace{2pc}%
\begin{minipage}{18pc}
\includegraphics[width=14pc,angle=270]{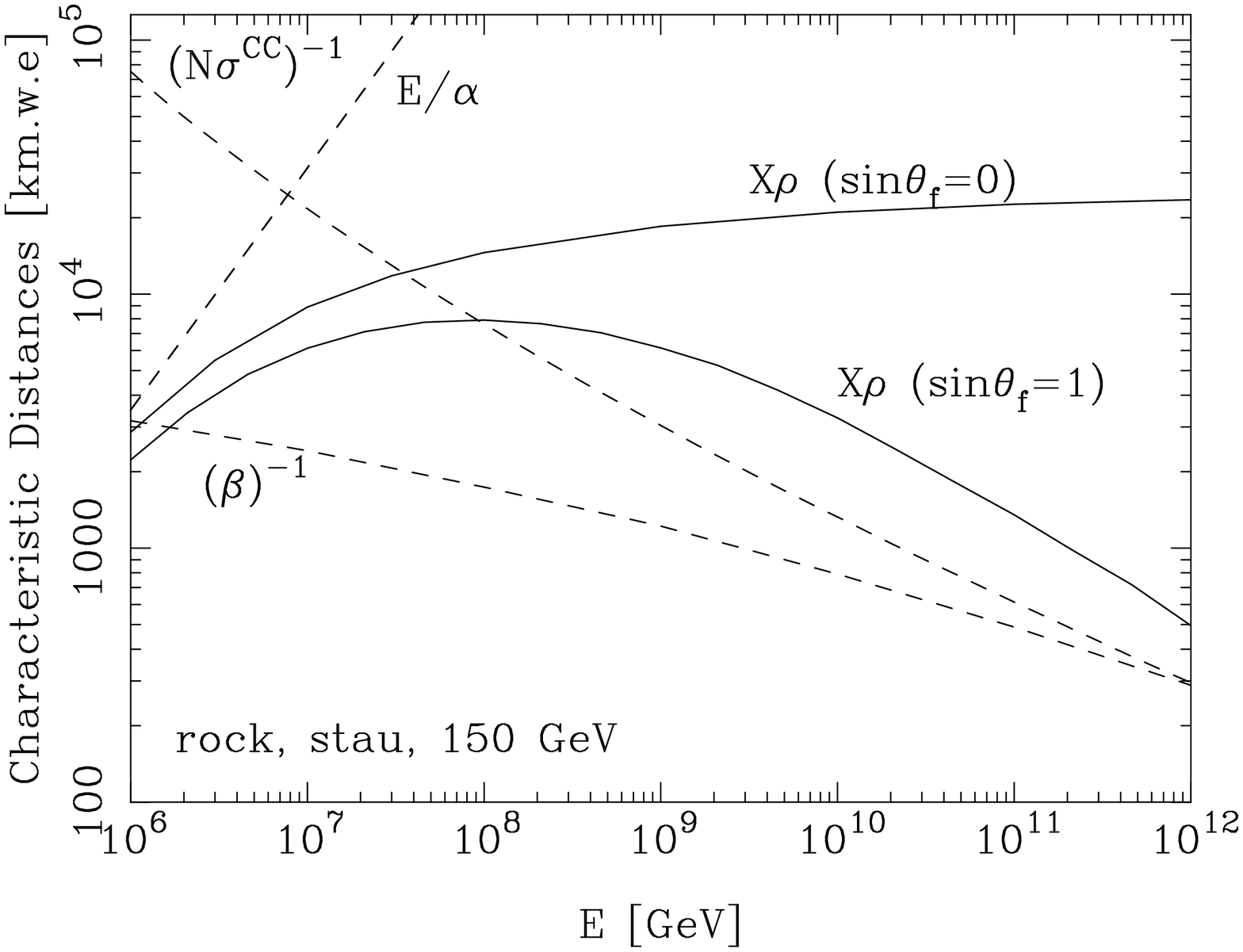}
\caption{\label{label}Characteristic distances 
in km.w.e.
for a stau with mass of
150 GeV moving through rock. Here $\sqrt{F}=10^7$ GeV in Eq. (3). The
upper solid line is the stau range with no CC interactions ($\sin\theta_f=0$), and the lower solid line is for maximal CC
interactions.}
\end{minipage} 
\end{figure}

\section{Conclusions}

We have shown here that the characteristic scales do not give an
accurate picture of the range of staus or taus. For taus, the characteristic
scale of the CC interaction length does show that the tau range is mainly
unaffected by CC interactions, even at $E=10^{12}$ GeV, as shown in
Fig. 3. The stochastic treatment of the staus is essential, however, for a determination
of the stau range. Because of electromagnetic energy loss, the stau
range is not equal to the CC interaction length.
Even when $\sin\theta_f=1$, the range is larger than
$L_{\tilde{\tau}}^{CC}$ at high energies because electromagnetic
interactions degrade the stau energy, and $L_{\tilde{\tau}}^{CC}$
increases with decreasing energy.

The results of Fig. 4 show that generically for neutrino
telescopes sensitive to fluxes below $10^6$ GeV, maximal
weak interactions of the staus will have a small numerical impact
on the event rate. On the other hand, experiments sensitive
to high energies like ANITA and its successors will be greatly
impacted by CC interactions if $\sin\theta_f\simeq 1$ given
measurable neutrino fluxes. The high energy signal of staus is
currently under investigation [13].

\ack
We thank J. Beacom for bring to our attention to the similarity in
distance scales for staus. 
We acknowledge Shufang Su's contributions to the work on electromagnetic energy loss of
staus and Yiwen Huang's contribution to the work on the weak interaction
cross section of staus. This work was supported in part by
DOE contracts DE-FG02-91ER40664, DE-FG02-04ER41319 and DE-FG02-04ER41298 
(Task C).


\section*{References}
\begin{thebibliography}{99}
\bibitem{doublebang}
J. Learned and S. Pakvasa 1995 {\it Astropart. Phys.}
{\bf 3} 267--274
\bibitem{neutrinotalk}
D. Cowen and T. deYoung, these proceedings
\bibitem{models}
For a review, see G. F. Giudice and R. Rattazzi 1999 {\it Phys. Rep.}
{\bf 332} 419-499
\bibitem{albuquerque}
I. Albuquerque, G. Burdman and Z. Chacko 2004 {\it Phys. Rev. Lett.}
{\bf 92} 221802
\bibitem{albuquerque2}
I. Albuquerque, G. Burdman and Z. Chacko 2006 eprint archive hep-ph/0605120
\bibitem{ringwald} 
M. Ahlers, J. Kersten and A. Ringwald 2006 {\it JCAP}
{\bf 0607} 005
\bibitem{rss}
M. H. Reno, I. Sarcevic and S. Su 2005 {\it Astropart. Phys.}
{\bf 24} 107--115
\bibitem{lep}
A. Heister {\it et al.} 2002 {\it Eur. Phys. J.} {\bf C 25} 339-351;
J. Abdallah {\it et al.} 2003 {\it Eur. Phys. J.} {\bf C 27} 153-172;
G. Abbiendi {\it et al.} 2003 {\it Phys. Lett. } B {\bf  572} 8-20
\bibitem{hrsu}
Y. Huang, M. H. Reno, I. Sarcevic and J. Uscinski 2006 eprint archive
hep-ph/0607216
\bibitem{fargion}
D. Fargion 2002 {\it Astrophys. J.} {\bf 570} 909-925; D. Fargion,
P. G. De Sanctis Lucentini, M. De Santis and M. Grossi 2004 {\it
Astrophys. J.} {\bf 613} 1285-1301
\bibitem{drss}
S. Iyer Dutta, M. H. Reno, I. Sarcevic and D. Seckel
2001 {\it Phys. Rev. }D {\bf 63} 094020
\bibitem{bugaev}
See also E. V. Bugaev and Y. V. Shlepin 2003
{\it Phys. Rev. }D {\bf 67} 034027
\bibitem{new}
I. Albuquerque, G. Burdman, M. H. Reno, I. Sarcevic, J.
Uscinski and Z. Chacko, in preparation

\end{thebibliography}
\end{document}